\documentclass[epj]{svjour}
\usepackage{graphicx,xcolor}
\usepackage{amsmath, amssymb, bm}
\usepackage{tikz}
\usepackage{comment}
\usepackage{ulem}
\begin{document}
\title{An approach to quasinormal modes of black hole based on reversed harmonic oscillator dynamics}
\author{Shigefumi Naka\inst{1} \and Haruki Toyoda\inst{2}
} 
%
%
\institute{Department of Physics, College of Science and Technology, Nihon University, Tokyo 101-8308, Japan
\and
Junior College, Funabashi Campus, Nihon University, Chiba 274-8501, Japan
}
\date{Received: date / Revised version: date}
%
\abstract{
The frequencies of quasinormal modes (QNM) for the Schwartzschild black hole are studied from the viewpoint of the particle scattering under an effective Regge-Wheeler type of potential consisting of a parabolic type one in an intermediate region and flat potentials on both sides. In particular, we use the eigenstates for a reversed harmonic oscillator as the complete bases in this intermediate region. Under this setting, the transmission and reflection coefficients are studied in addition to the frequencies of QNMs.
\PACS{
{03.65.-w}{Quantum Mechanics} \and
{03.65.Xp}{tunneling} \and
{04.70.Dy}{Quantum aspect of black holes, thermodynamics}
} 
} 
\maketitle
\section{Introduction}
\label{intro}

The quasinormal modes (QNM) for the Schwarzschild black hole\,(BH) are those of dynamical system describing the perturbation around the background black-hole spacetime \cite{RW},\cite{Zerilli},\cite{QNM-1},\cite{QNM-2}. Through the analysis of the QNM, the specific structures of those modes became evident; for example, the frequency characterizing scalar QNM have specific structure such as
\begin{align}
\omega_n=\omega\left\{\frac{1}{2\pi}\log 3 -i\left(n+\frac{1}{2}\right)+O(n^{-\frac{1}{2}}) \right\} \label{frequency}
\end{align}
($n \gg 1$), where $\omega$ is a constant determined by the surface gravity of the black holes. The frequency (\ref{frequency}) consists of two parts; that is, the equally spaced imaginary part, the $\Im\omega_n$, and the real part, the $\Re\omega_n$, which is proportional to $\log 3$ for a large quantum number $n$. The frequencies of the QNM were studied extensively from both analytic \cite{Motl}, \cite{Fiziev} and numerical ways\cite{Nollert}, since those are significant to make clear the stability and other properties of black holes.

The imaginary part of frequency was recognized since the early stage of the investigation of the QNM; for example, the WKB approximation for a tunneling potential characterizing the QNM gives bound state solutions with those imaginary frequencies\cite{Schutz-Will}, \cite{RUBAN}. The WKB approximation extracts an illustrative type of potential out of the QNM tunneling potential, and the essential reason for those imaginary frequencies exists at this point.

On the other hand, the factor $\log 3$ in the $\Re\omega_n$ was found initially by numerical methods, and it was later derived by analytical method, which is based on the monodromy of the exact solutions for a bound state equation associated with a radial coordinate\cite{Neitzke},\cite{Motl-2}. It is also an interesting point of view that the $\log 3$ is related to the area-spacing of a black hole\cite{Hod}, \cite{Dreyer},\cite{Setare}.

As for the real part of the $\omega_n$, studying its nature through various approaches is still significant. This paper aims to study a practical way of understanding the structure of the frequency $\omega_n$ through a scattering problem associated with the QNM.

The wave equation for the QNM can be reduced to the Schr$\ddot{\rm o}$dinger equation with a potential being partially parabolic in the one-dimensional space of the tortoise coordinate. In a previous paper\cite{RHO-PRD}, meanwhile, we showed that the Hamiltonian of reversed harmonic oscillator (RHO), a dynamical system with a parabolic potential, had taken the spectrum corresponding to the $\{\Im\omega_n\}$. Keeping this feature of RHO in mind, we investigate the scattering problem under a simplified setting such that the potential consists of parabolic and constant potentials on both sides with differences in level.

In the next section, we briefly review the previous paper on RHO in relation to the reduced wave equation of QNM. Therein, the $T$-representation of Green's function for the Sch$\ddot{\rm o}$dinger equation of RHO is discussed to make clear the meaning of the complete basis for RHO.

In Section 3, we develop the scattering theory under the simplified setting of the potential. The formulation is based on the standard quantum mechanics\cite{Ferrari-Mashhoon},\cite{Choudhury}; reflecting the shape of the potential, the region of the wave functions is divided into three parts, the I, I\!I, and I\!I\!I. In the regions I and I\!I\!I, the potential takes constant values, on which the wave functions are plane waves in the tortoise coordinate. Then, we set the wave functions in the region I\!I, as linear combinations of the complete bases for RHO. Under those settings, the transmission and reflection coefficients, the $|\mathcal{T}|^2$ and $|\mathcal{R}|^2$ respectively, are defined through the matching of respective wave functions at the boundaries of three regions\cite{Kemble}.

Section 4 is intended to derive the representations of the transmission and reflection coefficients and the energy spectrum $\{E\}$ of the particles, the counterpart of $\{\omega_n\}$, in terms of the wave functions in this scattering problem. Therefore, particle number conservation is essential to clear the energy spectrum. In this paper, the difference in the potential levels in the regions I and I\!I\!I is introduced but is dealt to be small. Because of this reason, the formulae in section 3 are reconstructed in section 4 by using the logarithmic derivatives of wave functions instead of wave functions themselves in the region I\!I. The discussion is also made on a Hawking-like temperature associated with the $|\mathcal{T}|^2$ given in the formalism.

In section 5, the relation between the transmission coefficient $|\mathcal{T}|^2$ and $\Re\tilde{E}$ is discussed based on the particle number conservation; in particular, we give attention to that relation at the neighborhood of the stationary point of the transmission coefficient. With the view of getting concrete results on $(\Re\tilde{E},|\mathcal{T}|^2)$ relation, we use two approximations to the wave functions, the asymptotic and semi-classical approximations. Under those approximation, the $\Re\tilde{E}$, which should be $\{\omega_n\}$ in the Eq.(\ref{frequency}), is discussed again from our point of view. Therein, a new viewpoint on the leading $\log$ term in the Eq.(\ref{frequency}) is also discussed in connection with the scale dependence of the $\Re\tilde{E}$.

Section 6 is devoted to summary and discussion. The appendices discuss the parametrization of the potential characterizing the QNM and other technical supplements to the text.

\section{Reversed harmonic oscillator in the QNM}
\label{sec:1}

\begin{figure}
\center
\begin{minipage}{5cm}
\includegraphics[width=5cm]{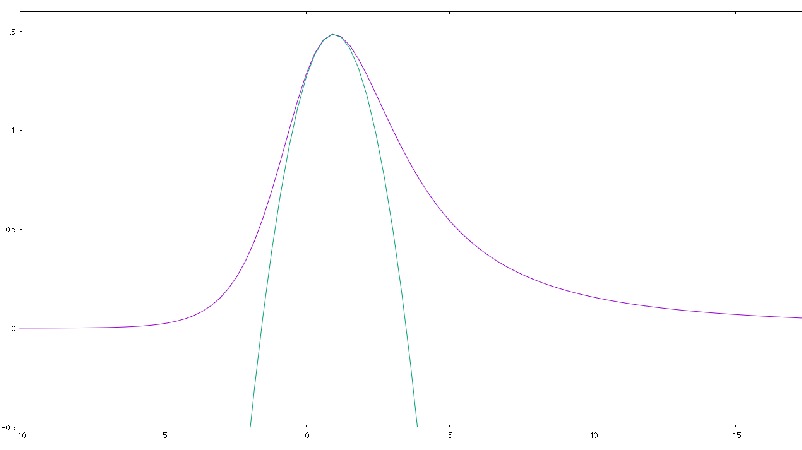}
\end{minipage}
\caption{The real line and dashed line are respectively $V_{RW}$ and its parabolic approximation having their maximal points in common. The horizontal line designates $\tilde{x}$ . }
\label{fig:1}
\end{figure}

The spacetime for a mass $M$ Schwarzschild BH is characterized by the metric $\mathring{g}_{\mu\nu}$ defined by
\begin{align}
ds^2 &=-f(r)(dx^0)^2+\frac{1}{f(r)}dr^2+r^2d\Omega_{(2)}^2 , \label{Sch-BH}
\end{align}
where $dx^0=cdt$, and the $d\Omega_{(2)}^2$ is the line element on the unit 2-sphere, and
\begin{align}
f(r) &=1-\frac{r_H}{r}.
\end{align}
Here, we have written the horizon's radius as $r_H=\frac{2MG_4}{c^2}$ with the gravitational constant $G_4$ in 4-dimensional spacetime. The QNM are radial components $\{Q(h_{\mu\nu})\}$ of a perturbation of the background metric $\mathring{g}_{\mu\nu}+h_{\mu\nu}$. Those components are characterized by the wave equation outside the event horizon such that
\begin{align}
\left[ -(\partial_0)^2+(\partial_x)^2-V_{RW}\big(r(x)\big) \right]Q=0, \label{QNM}
\end{align}
where $V_{RW}$ is the Regge-Wheeler potential\cite{RW},\cite{Zerilli}
\begin{align}
V_{RW}(r)=f(r)\left\{\frac{a_L}{r^2}-\frac{r_Hb_J}{r^3}\right\}
\end{align}
; and, $x(r)=r+r_H\ln \left|\frac{r}{r_H}-1\right|$ is the tortoise coordinate, which maps the region $r_H\leq r<\infty$ to $-\infty<x<\infty$. The notations such as $a_L:=L(L+1)\,(L\in\mathbb{N})$ and $b_J\,(J=0,1,2)$ are also used to designate respectively the eigenvalues of $\Delta_2$ and the spin of the perturbing field $Q$. Then, for example for $(L,J)=(2,2)$, the shape of dimensionless potential $\tilde{V}_{RW}=r_H^2V_{RW}$ as a function of dimensionless coordinate $\tilde{x}=r_H^{-1}x$ becomes the real line in Fig.\ref{fig:1}.

In the case of $Q(t,x)=\Psi(x)e^{-\frac{i}{\hbar}\epsilon t}$, with a mass dimensional constant $m_*=\frac{\hbar}{r_Hc}$ introduced for descriptive purpose, Eq.(\ref{QNM}) is expressed as a one dimensional time-independent Schr$\ddot{\rm o}$dinger equation
\begin{align}
\hat{H}_{RW}\Psi=E_{RW}\Psi~~\left(\, \hat{H}_{RW}=\frac{1}{2m_*}\hat{p}_x^2+V(x) \,\right) \label{Sch-RW}
\end{align}
, where $V(x)=\frac{\hbar^2}{2m_*}V_{RW}(x)$ and $E_{RW}=\frac{1}{2m_*}\left(\frac{\epsilon}{c}\right)^2$.

The Eq.(\ref{Sch-RW}) is a second-order differential equation concerning $x$, and there exist exact solutions in terms of the confluent Heun's functions\cite{exact}. To know the physical properties of the QNM based on Eq.(\ref{QNM}), it is rather required to study the eigenvalue of $\hat{H}_{RW}$, the S-matrix by $\hat{H}_{RW}$, and so on. For this purpose, it is useful to approximate $V_{RW}$ by some potentials, by which those problems come to be solvable. One of the good approximations in this sense is a parabolic potential, the dashed curve in Fig.\ref{fig:1}, inscribed in $V_{RW}(x)$, with the maximum point $x_0$ common to $V_{RW}(x)$. Here, writing two roots of $\partial_{\tilde{x}}V_{RW}(x)=0$ as $r_{+}>r_{-}$, the maximum point becomes $x_0=x(r_{+})$ due to $r_{-}<r_H$ (Appendix A). Further, since $\partial^2_{\tilde{x}}V_{RW}(x_0)<0$, one can approximate the potential $V(x)$ by
\begin{align}
V(x)\simeq V_K(x)\equiv V_0-\frac{m_*\omega_*^2}{2}(x-x_0)^2, \label{parabolic}
\end{align}
where $V_0=V(x_0)$ and $\omega_*=\sqrt{\frac{1}{m_*}\left|\partial_x^2V\right|_0}$.
The tunneling problem through the parabolic potential (\ref{parabolic}) was studied first by E. Kemble\cite{Kemble} in his WKB method. This WKB approach was also useful for analyzing QNM\cite{Schutz-Will}.
Writing $x$ in the sense of $x-x_0$ anew, the eigenvalue problem of $\hat{H}_{RW}$ can be solved approximately by studying the Hamiltonian
\begin{align}
\hat{H}_K=\frac{1}{2m_*}\hat{p}_x^2+V_K(x)=\hat{H}_r+V_0 \label{H_K},
\end{align}
where
\begin{align}
\hat{H}_r=\frac{1}{2m_*}\hat{p}_x^2-\frac{m_*\omega_*}{2}x^2 \label{H_r}
\end{align}
is a Hamiltonian for a RHO. The eigenvalues problem of $\hat{H}_r$ can be solved easily with the aid of the ladder operators\cite{RHO-PRD}
\begin{align}
\begin{split}
A &=\sqrt{\frac{m_*\omega_*}{2\hbar}}x-\frac{1}{\sqrt{2m_*\hbar\omega_*}}\hat{p}_x , \\
\bar{A} &=\sqrt{\frac{m_*\omega_*}{2\hbar}}x+\frac{1}{\sqrt{2m_*\hbar\omega_*}}\hat{p}_x \,,
\end{split} \label{ladder}
\end{align}
to which $A=A^\dag,\bar{A}=\bar{A}^\dag$ and $[A,\bar{A}]=-[\bar{A},A]=i$ hold. In terms of those ladder operators the Hamiltonian (\ref{H_r}) can be written as $\hat{H}_r=-i\hbar\omega_*\left(\Lambda+\frac{1}{2}\right)=-i\hbar\omega_*\left(\bar{\Lambda}-\frac{1}{2}\right)$, where $\Lambda=-i\bar{A}A$ and $\bar{\Lambda}=-iA\bar{A}(=\Lambda+1)$. Since $[\Lambda,\bar{A}]=\bar{A}$ and $[\bar{\Lambda},A]=-A$, one can say that $A$ and $\bar{A}$ work as the ladder operators for $\Lambda$ and $\bar{\Lambda}$ respectively; and, the states $\phi_{(0)}(x)=\bar{\phi}_{(0)}(x)^*=\sqrt[4]{\frac{m_*\omega_*}{2\hbar\pi^2}}e^{i\frac{m_*\omega_*}{2\hbar}x^2}$ play the role of cyclic states, the counterpart of ground state in a harmonic oscillator, characterized by $A\phi_{(0)}=\bar{A}\bar{\phi}_{(0)}=0$ .Then, the base states, the counterparts of excited states in the harmonic oscillator, are given by
\begin{align}
\begin{split}
\phi_{(n)}(x) &=\bar{A}^n\phi_{(0)}(x), \\
\bar{\phi}_{(n)}(x) &=A^n\bar{\phi}_{(0)}(x)
\end{split} \label{phi-n}
\end{align}
$(n=1,2,\cdots)$, to which one can verify $\hat{H}_r\phi_{(n)}=E_n\phi_{(n)}$ and $\hat{H}_r\bar{\phi}_{(n)}=-E_n\bar{\phi}_{(n)}$, where
\begin{align}
E_n=-i\hbar\omega_*\left(n+\frac{1}{2}\right)~(n=0,1,2,\cdots). \label{E-n}
\end{align}
This means that $\hat{H}_r\bar{\phi}_{(\bar{n})}=E_n\bar{\phi}_{(\bar{n})}$ for $\bar{n}=-(n+1)$ under analytic continuation of the suffixes $\{n\}$ of $\{\bar{\phi}_{(n)}\}$ to negative values $\{\bar{n}\}$; then, the independent states belonging to the same eigenvalue $E_n$ are $\{\phi_{(n)},\bar{\phi}_{(\bar{n})}\}$. By definition, the states $\{\phi_{(n)},\bar{\phi}_{(n)}\}$ satisfy the normalization $\langle \bar{\phi}_{(m)}|\phi_{(n)}\rangle=\delta_{m,n}N_n$ with $N_n=i^nn!\sqrt{\frac{i}{2\pi}},(n,m=0,1,2,\cdots)$.

Now, the importance is that $\bar{\phi}_{(n)}(x)=\phi_{(n)}^*(x)\,(n\in \mathbb{N})$ in the $x$-representation; and, $\{\phi_{(n)},\bar{\phi}_{(n)}\}$ form a complete basis\cite{RHO-PRD} in such a sense that
\begin{align}
1=\sum_{n=0}^\infty\frac{1}{N_n} |\phi_{(n)}\rangle\langle \bar{\phi}_{(n)}|=\sum_{n=0}^\infty\frac{1}{N_n^*} |\bar{\phi}_{(n)}\rangle\langle \phi_{(n)}|. \label{complete-1}
\end{align}
If we use Eq.(\ref{complete-1}) as a trial to represent the Green's function of the Sch$\ddot{\rm o}$dinger equation for the potential $V_K$, one can derive within the framework of the ladder operator formalism (appendix B) so that
\begin{align}
G(T &;x_a,x_b)=\langle x_b|e^{-\frac{i}{\hbar}\hat{H}_rT}|x_a\rangle \nonumber \\
&=\sqrt{\frac{m\omega_*}{2\pi i\hbar\sinh(\omega_* T)}} \nonumber \\
&\times e^{i\frac{m_*\omega_*}{2\hbar\sinh(\omega_* T)}\left\{\cosh(\omega_* T)\left(x_b^2+x_a^2\right)-2x_bx_a\right\}}. \label{Green-T2}
\end{align}
The result is a form of Green's function for a harmonic oscillator with the angular frequency $\omega$ followed by an analytic continuation $\omega\rightarrow i\omega_*$ as expected. The limit $T\rightarrow 0$ of Eq.(\ref{Green-T2}) leads to
\begin{align}
\langle x_b|x_a \rangle=\sum_{n=0}^\infty \frac{1}{N_n}\phi_{(n)}(x_a)\phi_{(n)}(x_b)=\delta(x_a-x_b) \label{complete-2}
\end{align}
, which means the validity of Eq.(\ref{complete-1}) in the form of matrix elements. The second equality of Eq.(\ref{complete-1}) is nothing but the complex conjugate of Eq.(\ref{complete-2}).
Until now, we have written the energy eigenvalue associated with the suffix $n$ of $\phi_{(n)}$ as $E_n$. It is, however, convenient to write suffixes of the state as $n_E$ associated with an energy eigenvalue $E$ of $\hat{H}_r$. In the following, we rather use the nations as $n_E\equiv i\tilde{E}-\frac{1}{2}\,\left(\tilde{E}=E/\hbar\omega_*\right)$ including the case of a complex $\tilde{E}$. Then using $n_E$ and $\bar{n}_E\equiv \left(n_{E^*}\right)^*=-i\tilde{E}-\frac{1}{2}=-(n_E-1)$ instead of $(n,\bar{n})$, the $\phi_{(n_E)}$ and $\bar{\phi}_{(\bar{n}_E)}=\left(\phi_{(n_{E^*})}\right)^*$ become independent states associated with a common energy $E$.
\section{Scattering problem by a truncated parabolic potential}

\begin{figure}
\center
\begin{minipage}{6cm}
\includegraphics[width=6cm]{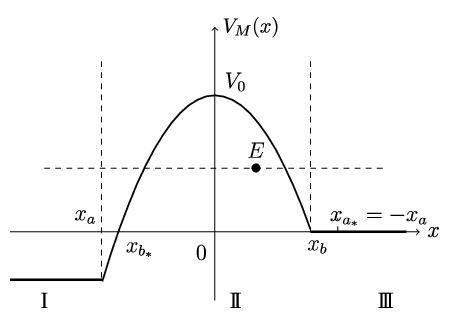}
\end{minipage}
\caption{($x_a<x_{b_*}=-x_b$).~
The $V_M(x)$ is the potential truncated from $V_K$(x) so that $V_M(x)=V_K(x)\theta(x_b-x)\theta(x-x_a)+V_K(x_a)\theta(x_a-x)$. The symbols of regions are I$=\{x\leq x_a\}$,\, I\!I=$\{x_a<x<x_b\}$ and I\!I\!I=$\{x_b\leq x\}$ respectively.
}
\label{fig:2}
\end{figure}

The parabolic potential $V_K(x)$ does not reflect the vanishing asymptotic behavior of $V_{RW}(x)$ as $|x|\rightarrow \infty$, though the eigenvalue problem of $\hat{H}_K$ can be solved exactly. Then, we consider a model such that the potential $V_K(x)$ is replaced by a modified potential $V_M(x)$ illustrated in Fig.\ref{fig:2}.
The Schr$\ddot{\rm o}$dinger equation for this dynamical system is
\begin{align}
\hat{H}_M\Psi=E\Psi~~\left(\, \hat{H}_M:=\frac{1}{2m_*}\hat{p}_x^2+V_M(x) \,\right). \label{modified}
\end{align}

Since the potential $V_M(x)$ is continuous and bounded, the wave functions and their derivative should be continuous at the linking points $x_a$ and $x_b$. Further, on account of that, the potential $V_M$ reduces respectively to $V_K(x_a)(=\mbox{const.})$ in the region I and $V_K(x_b)=0$ in the region I\!I\!I, we can put the corresponding wave functions in terms of dimensionless coordinate
\footnote{We use $l_*=\sqrt{\frac{\hbar}{m_*\omega_*}}$\, $\left(=r_H\sqrt[4]{2/|V^{\prime\prime}(0)|}\right)$ as a constant with the dimension of length. Further, throughout this paper, the \lq tilde \rq is used to represent dimensionless quantities such as $\tilde{x},\tilde{k},\tilde{E}$, et cetera. In terms of dimensionless variables, the potential of $H_K$ can be written as $\tilde{V}_K(\tilde{x})=-\frac{1}{2}\tilde{x}^2+\tilde{V}_0$.}
$\tilde{x}=l_*^{-1}x$ so that
\begin{align}
\psi_E^{({\rm I})}(x) &=Ae^{-i\tilde{k}^\prime(\tilde{x}-\tilde{x}_a)}, \\
\psi_E^{(\rm{I\!I\!I})}(x) &=C_{(+)}e^{i\tilde{k}(\tilde{x}-\tilde{x}_b)}+C_{(-)}e^{-i\tilde{k}(\tilde{x}-\tilde{x}_b)}
\end{align}
, where $(A,C_{(\pm)})$ are constants; $\tilde{k}=l_* k$ and $\tilde{k}^\prime=l_* k^\prime$ are dimensionless variables associating respectively with the incoming and outgoing wave numbers $k$ and $k^\prime$. Generally, we may regard the wave numbers as $k\neq k^\prime$ reflecting $x_a\neq x_{b_*}$. In the present situation, $\tilde{x}=-\infty,0$, and $\infty$ are corresponding respectively to the horizon $r_H$, the peak position of $V_{RW}$, and $r=\infty$. We are also considering the scattering problem with incident particles from the right side, and $\psi_E^{\rm (I)}(x)$ is regarded as an outgoing state. If we define the energy $\tilde{E}$ of an incident particle by $\tilde{k}=\sqrt{2\tilde{E}}$, then the Fig.2 leads to $\tilde{k}^\prime=\sqrt{2\left(\tilde{E}+\frac{1}{2}(\tilde{x}_a^2-\tilde{x}_b^2)\right)}$. This implies that for a small $\delta\equiv \tilde{x}_{a_*}-\tilde{x}_b\ll 1$, one can write $\Delta\tilde{k}\equiv \tilde{k}^\prime-\tilde{k}\simeq \frac{1}{2}\frac{\tilde{x}_{ab}}{\tilde{k}}\delta,\,\left(\tilde{x}_{ab}=\frac{1}{2}(\tilde{x}_{a_*}+\tilde{x}_b)\right)$ within the first order of $\delta$; that is, $\Delta\tilde{k}=O(\delta)$.

In the region I\!I, the wave function $\psi_E^{\rm (I\!I)}(x)$ is characterized by an eigenvalue equation $\mathcal{H}_K\psi_E^{\rm (I\!I)}(x)=E\psi_E^{\rm (I\!I)}(x)\,(x\in {\rm I\!I})$. Furthermore, the $E=\hbar\omega_*\tilde{E}$ may be a complex value in this region due to liking conditions of wave functions at boundaries and low particle number conservation. Under those backgrounds, we try to set
\begin{align}
\psi_E^{\rm (I\!I)}(x) &= B_1\phi_{(n_E)}(x)+B_2\bar{\phi}_{(\bar{n}_E)}(x)~~\left(x\in{\rm I\!I}\right)
\end{align}
, where $B_1,B_2$ are constants.

Under those setting of wave functions, the linking conditions at $x_a$ become
\begin{align}
A &=B_1\phi_{(n_E)a}+B_2\bar{\phi}_{(\bar{n}_E)a}, \\
-i\tilde{k}^\prime A &=B_1\phi_{(n_E)\mathring{a}}+B_2\bar{\phi}_{(\bar{n}_E)\mathring{a}}
\end{align}
, where henceforth, we use the abbreviation such as $\phi_{(n_E)a}=\phi(x_a),\,\phi_{(n_E)\mathring{a}}=\partial_{\tilde{x}_a}\phi_{(n_E)}(x_a)$.
Similarly, the linking conditions at $x_b$ become
\begin{align}
B_1\phi_{(n_E)a}+B_2\bar{\phi}_{(\bar{n}_E)b} &=C_{(+)}+C_{(-)},\\
B_1\phi_{(n_E)\mathring{b}}(x_b)+B_2\bar{\phi}_{(\bar{n}_E)\mathring{b}} &=i\tilde{k}\left(C_{(+)}-C_{(-)}\right).
\end{align}
Out of those linking equations, the $B_i(i=1,2)$ can be eliminated easily, and we have in matrix form that
\begin{align}
A\begin{bmatrix}1 \\ -i\tilde{k}^\prime \end{bmatrix} &=\frac{1}{\Phi_{b\mathring{b}}}
\begin{bmatrix} \Phi_{a\mathring{b}}-i\tilde{k}\Phi_{ab} & \Phi_{a\mathring{b}} +i\tilde{k}\Phi_{ab} \\
\phi_{\mathring{a}\mathring{b}}-i\tilde{k}\Phi_{\mathring{a}b} & \Phi_{\mathring{a}\mathring{b}}+i\tilde{k}\Phi_{\mathring{a}b}
\end{bmatrix}
\begin{bmatrix} C_{(+)} \\ C_{(-)} \end{bmatrix}
\end{align}
, where $\Phi_{b\mathring{b}}=\phi_{(n_E)b}\bar{\phi}_{(\bar{n}_E)\mathring{b}}-\bar{\phi}_{(\bar{n}_E)\mathring{b}}\phi_{(n_E)b}$, and
\begin{align}
\Phi_{ab} &=\phi_{(n_E)a}\bar{\phi}_{(\bar{n}_E)b}-\bar{\phi}_{(\bar{n}_E)a}\phi_{(n_E)b}, \\
\Phi_{a\mathring{b}} &=\phi_{(n_E)a}\bar{\phi}_{(\bar{n}_E)\mathring{b}}-\bar{\phi}_{(\bar{n}_E)a}\phi_{(n_E)\mathring{b}}, \\
\Phi_{\mathring{a}b} &=\phi_{(n_E)\mathring{a}}\bar{\phi}_{(\bar{n}_E)b}-\bar{\phi}_{(\bar{n}_E)\mathring{a}}\phi_{(n_E)b}, \\
\Phi_{\mathring{a}\mathring{b}} &=\phi_{(n_E)\mathring{a}}\bar{\phi}_{(\bar{n}_E)\mathring{b}}-\bar{\phi}_{(\bar{n}_E)\mathring{a}}\phi_{(n_E)\mathring{b}}.
\end{align}
Then, defining $\mathcal{T}=\frac{A}{C_{(-)}}$ and $\mathcal{R}=\frac{C_{(+)}}{C_{(-)}}$, the linking conditions turn out to be
\begin{align}
\mathcal{T} &=\left(\tilde{\Phi}_{a\mathring{b}}-i\tilde{k}\tilde{\Phi}_{ab}\right)\mathcal{R}+\tilde{\Phi}_{a\mathring{b}} +i\tilde{k}\tilde{\Phi}_{ab}, \\
-i\tilde{k}^\prime\mathcal{T} &=\left(\tilde{\Phi}_{\mathring{a}\mathring{b}}-i\tilde{k}\tilde{\Phi}_{\mathring{a}b}\right)\mathcal{R}+\tilde{\Phi}_{\mathring{a}\mathring{b}}+i\tilde{k}\tilde{\Phi}_{\mathring{a}b}
\end{align}
, where $\tilde{\Phi}_{ab}=\Phi_{ab}/\Phi_{b\mathring{b}},\, \tilde{\Phi}_{a\mathring{b}}=\Phi_{a\mathring{b}}/\Phi_{b\mathring{b}}$ and so on. The $\mathcal{T}$ and $\mathcal{R}$ are respectively densities of transmission coefficient and reflection coefficient, which can be solved as
\begin{align}
\mathcal{T} &=2i\tilde{k}\frac{\left(\tilde{\Phi}_{\mathring{a}\mathring{b}}\tilde{\Phi}_{ab}-\tilde{\Phi}_{\mathring{a}b}\tilde{\Phi}_{a\mathring{b}}\right)}{\left(\tilde{\Phi}_{\mathring{a}\mathring{b}}-i\tilde{k}\tilde{\Phi}_{\mathring{a}b}\right)+i\tilde{k}^\prime\left(\tilde{\Phi}_{a\mathring{b}} -i\tilde{k}\tilde{\Phi}_{ab}\right)} , \label{T-eq} \\
\mathcal{R} &=-\frac{i\tilde{k}^\prime\left(\tilde{\Phi}_{a\mathring{b}}+i\tilde{k}\tilde{\Phi}_{ab}\right)+\left(\tilde{\Phi}_{\mathring{a}\mathring{b}}+i\tilde{k}\tilde{\Phi}_{\mathring{a}b}\right)}
{i\tilde{k}^\prime\left(\tilde{\Phi}_{a\mathring{b}}-i\tilde{k}\tilde{\Phi}_{ab}\right)+\left(\tilde{\Phi}_{\mathring{a}\mathring{b}}-i\tilde{k}\tilde{\Phi}_{\mathring{a}b}\right)}. \label{R-eq}
\end{align}
It should be noticed that those $\mathcal{T}$ and $\mathcal{R}$ have the structures such as $\mathcal{T}=2i\tilde{k}\frac{C(k)}{D(k,k^\prime)}$ and $\mathcal{R}=-\frac{D(-k,k^\prime)}{D(k,k^\prime)}$, where $C(k)=\left(\tilde{\Phi}_{\mathring{a}\mathring{b}}\tilde{\Phi}_{ab}-\tilde{\Phi}_{\mathring{a}b}\tilde{\Phi}_{a\mathring{b}}\right)$ and $D(k,k^\prime)=i\tilde{k}^\prime\left(\tilde{\Phi}_{a\mathring{b}}-i\tilde{k}\Phi_{ab}\right)$ $+\left(\tilde{\Phi}_{\mathring{a}\mathring{b}}-i\tilde{k}\tilde{\Phi}_{\mathring{a}b}\right)$.

The conservation of particle number is represented in terms of the transmission coefficient $|\mathcal{T}|^2$ and the reflection coefficient $|\mathcal{R}|^2$ so that
\begin{align}
\left|T\right|^2+\left|R\right|^2=1. \label{conservation}
\end{align}
We note that the $\tilde{E}$ is a consistent solution of linking conditions and Eq.(\ref{conservation}), which becomes a complex value depending on $(\tilde{k},\tilde{k}^\prime)$ in general.

\section{$(\mathcal{T},\mathcal{R})$ in terms of derivatives of $(\phi,\bar{\phi})$}

To see a more aspect of $(\mathcal{T},\mathcal{R})$, it is useful to introduce the logarithmic derivatives $(K,\bar{K})$ of $(\phi,\bar{\phi})$ defined by
\begin{align}
\phi_{(n_E)\mathring{c}} &=iK_{(n_E)c}\phi_{(n_E)c}, \\
\bar{\phi}_{(\bar{n}_E)\mathring{c}} &=-i\bar{K}_{(\bar{n}_E)c}\bar{\phi}_{(\bar{n}_E)c}
\end{align}
for any $\tilde{x}_c$; then $\bar{\phi}_{(\bar{n}_E)\mathring{c}}=\left(\phi_{(n_{E^*})\mathring{c}}\right)^*$$=\left(iK_{(n_{E^*})c}\phi_{(n_{E^*})c}\right)^*=-iK_{(\bar{n}_E)c}^*\bar{\phi}_{(\bar{n}_E)c}$ by definition of $\bar{\phi}$ and $\bar{n}_E\equiv \left(n_{E^*}\right)^*=-(n_E+1)$. This equation implies $\bar{K}_{(\bar{n}_E)c}=\left(K_{(n_{E^*})c}\right)^*$.

We, further, consider $\{\tilde{x}\}$ up to this stage as the coordinate on the real axis in a complex $\{\tilde{x}\}$-plane. Since the $\phi_{(n_E)}(x)$ is analytic in a region aside from $\tilde{x}=0$, $\phi_{(n_E)}(x_{a_*})$ and $\phi_{(n_E)}(x_a)$ is related by analytic continuation going round a semicircle in upper half of the complex-plane $\tilde{x}_{a_*}\rightarrow e^{i\pi}\tilde{x}_{a_*}(=\tilde{x}_a)$; and, the Eq.(\ref{phi-n}) yields
\begin{align}
\phi_{(n_E)a}=e^{i\pi n_E}\phi_{(n_E)a_*} , \label{a-a_*} \\
\bar{\phi}_{(\bar{n}_E)a}=e^{-i\pi \bar{n}_E}\bar{\phi}_{(\bar{n}_E)a_*}.
\end{align}
Conversely, we must be careful about the derivative of those equations. As in Appendix(C), we obtain
\begin{align}
\phi_{(n_E)\mathring{a}} &=-e^{i\pi n_E}iK_{(n_E)a_*}\phi_{(n_E)a_*}, \label{K-dot}\\
\bar{\phi}_{(n_E)\mathring{a}} &=-e^{i\pi \bar{n}_E}i\bar{K}_{(\bar{n}_E)a_*}\bar{\phi}_{(\bar{n}_E)a_*}. \label{bar-K-dot}
\end{align}
The forms of $(\ref{T-eq})$ and $(\ref{R-eq})$ in terms of $(K,\bar{K})$ are still complex; however, in the case of a small interval $\delta \equiv \tilde{x}_{a_*}-\tilde{x}_b\ll 1$, which we assume, the representations of $C(k)$ and $D(k,k^\prime)$ become rather simple within the first order of $\delta$ so that
\begin{align}
C(k) &\sim -\left(\frac{e^{i\pi n_E}}{N_{b\mathring{b}}}\right)^2\left\{\left(K+\bar{K}\right)^2+A_\delta \right\}, \\
D(k,k^\prime) &\sim 2\left(\frac{e^{i\pi n_E}}{N_{b\mathring{b}}}\right)\Bigg\{\left(\eta-K\right)\left(\eta+\bar{K}\right) \nonumber \\
&\hspace{20mm} -\left(\frac{\Delta\tilde{k}}{2}\right)^2+B_\Delta\Bigg\} \label{Dkk}
\end{align}
, where $N_{b\mathring{b}}=-i\left(K+\bar{K}\right)$, $\eta=\frac{1}{2}\left(\tilde{k}^\prime+\tilde{k}\right)$, $\Delta\tilde{k}=\tilde{k}^\prime -\tilde{k}$ with the abbreviation $\left(K_{(n_E)b},\bar{K}_{(\bar{n}_E)b}\right)\rightarrow \left(K,\bar{K}\right)$. In addition to those, we have used the notations
\begin{align}
A_\delta &=i\delta\left(K-\bar{K}\right)^3-\delta\left(K-\bar{K}\right)^2, \label{A-delta}\\
B_\Delta &=-\left(\frac{\Delta\tilde{k}}{2}\right)^2+i\delta \frac{1}{2}\left(K+\bar{K}\right)^2+\delta\frac{1}{2}\left(K-\bar{K}\right). \label{B-Delta}
\end{align}
Here, the first term in the right-hand side of Eq.(\ref{B-Delta}) should be read as 0 because of $\Delta\tilde{k}=O(\delta)$; and so, if necessary, we may regard $B_\Delta=O(\delta)$. However, the form of Eq.(\ref{B-Delta}) is useful to derive $D(-k,k^\prime)$ through $D(-k,k^\prime)\equiv \left. D(k,k^\prime)\right|_{\eta\rightarrow 0, \Delta\tilde{k}\rightarrow -2\eta}$.

Thus substituting the resultant $C(k),~ D(k,k^\prime)$, and $D(-k,k^\prime)$ for $(\mathcal{T},\mathcal{R})$, we arrive at
\begin{align}
\mathcal{T} &\sim -i\tilde{k}\left(\frac{e^{i\pi n_E}}{N_{b\mathring{b}}}\right)\frac{\left(K+\bar{K}\right)^2+A_\delta}{\left(\eta-K\right)\left(\eta+\bar{K}\right)+B_\Delta}, \label{AT-eq} \\
\mathcal{R} &\sim \frac{\frac{\Delta\tilde{k}}{2}\left(K-\bar{K}\right)-K\bar{K}+\eta^2-B_\Delta}{\left(\eta-K\right)\left(\eta+\bar{K}\right)+B_\Delta}. \label{AR-eq}
\end{align}
The transmission coefficient $\left|\mathcal{T}\right|^2$ defined through Eq.(\ref{AT-eq}) has a characteristic form
\footnote{
The $|\mathcal{T}|^2$ in Eq.(\ref{transmission}) includes the transmission coefficient for the rectangular potential $V_M=V_0=\mbox{const.}\,(>0)$ with $\delta=0$. Indeed, removing first the factor $e^{-2\pi\Re\tilde{E}}$ by $\mathcal{T}\rightarrow e^{i\pi n}\mathcal{T}$, and putting secondly $K+\bar{K}=(2i\kappa)\sinh^{-1}(\kappa\bar{x})$, $K-\bar{K}=(2i\kappa)\cosh(\kappa\bar{x})\sinh^{-1}(\kappa\bar{x})$ with $\eta=\tilde{k}=\sqrt{2\tilde{E}}$ and $\kappa=\sqrt{2(\tilde{V}_0-\tilde{E})}$, we obtain \vspace{-2mm}
\begin{align*}
|\mathcal{T}|^2\sim\left|\frac{\tilde{k}}{\tilde{k}\cosh(2\kappa\bar{x})+\frac{i}{2\kappa}(\kappa^2-\tilde{k}^2)\sinh(2\kappa\bar{x})}\right|^2
\end{align*}
, where $\bar{x}=\tilde{x}_a+\tilde{x}_b$. The result is the transmission coefficient's standard form, which satisfies $|\mathcal{T}|^2\rightarrow 1~(E\rightarrow \infty)$.
}
; because of $\Im{n}_E=\Re\tilde{E}$,
\begin{align}
\left|\mathcal{T}\right|^2 &\sim e^{-2\pi\Re\tilde{E}}\left|\tilde{k}\frac{\left(K+\bar{K}\right)+\left(K+\bar{K}\right)^{-1}A_\delta}{\left(\eta-K\right)\left(\eta+\bar{K}\right)+B_\Delta}\right|^2. \label{transmission}
\end{align}
The right-hand side of this equation has a stationary point as a function of $\eta$, and the transmission coefficient behaves as $\left|\mathcal{T}\right|^2\sim \mbox{const.}\times e^{-2\pi\Re\tilde{E}}$ in a neighborhood of such a point. Since $\Re\tilde{E}$ is the energy of the particles in the region ${\rm I\!I}$, it is interesting to regard $e^{-2\pi\Re\tilde{E}}$ as a thermal-like energy distribution for the particles in this region. Then putting $\pi\Re\tilde{E}=\beta_*\Re E$ $\left(\beta_*=\frac{1}{k_BT_*}\right)$, we obtain
\begin{align} T_*=\frac{\hbar\omega_*}{2\pi k_B}=\frac{\hbar c}{2\pi k_B r_H}\sqrt{\frac{1}{2}\left|\partial^2_{\tilde{x}}\tilde{V}_{RW}\right|_0}~.
\end{align}
The $T_*$ is nothing but the Hawking temperature multiplied by an order one factor $\sqrt{\frac{1}{2}\left|\partial^2_{\tilde{x}}\tilde{V}_{RW}\right|_0}$.

\section{The role of the particle number conservation}

Using the Eq.(\ref{AT-eq}) and Eq.(\ref{AR-eq}), the particle number conservation (\ref{conservation}) gives another representation of $|\mathcal{T}|^2$ so that
\begin{align}
\left|\mathcal{T}\right|^2\sim 1-\left|1+f\left(\eta\right)\right|^2
\label{T-PNC}
\end{align}
, where $\eta\equiv \frac{1}{2}\left(\tilde{k}^\prime+\tilde{k}\right)$ and
\begin{align}
f(\eta) &=\frac{\left(\eta+\frac{\Delta\tilde{k}}{2}\right)(K-\bar{K})-2B_\Delta}{\eta^2-K\bar{K}-\eta\left(K-\bar{K}\right)+B_\Delta}. \label{T-f}
\end{align}
Equating the representations of $\left|\mathcal{T}\right|^2$ in Eq.(\ref{transmission}) and Eq.(\ref{T-PNC}), and solving its with respect to $\Re\tilde{E}$, we obtain
\begin{align}
\Re\tilde{E} &\sim \frac{1}{2\pi}\log \frac{\left(\eta-\frac{\Delta\tilde{k}}{2}\right)^2\left|(K+\bar{K})^2+2A_\delta\right|}{\left|\mathcal{T}\right|^2\left|\left(\eta-K\right)\left(\eta+\bar{K}\right)+B_\Delta \right|^2} \label{ReE-1}
\end{align}
, where the terms of $O(\delta^2)$ are disregarded.

The stationary point $\eta_s$ of $|\mathcal{T}|^2$ as a function of $\eta$ is defined by $\frac{\partial}{\partial\eta}|\mathcal{T}|^2=0$. From the Eq.(\ref{T-PNC}), one can find that the $\eta_s$ characterized by $|f(\eta_s)+1|=0$ or by
\begin{align}
\eta_s^2-\left(K\bar{K}-\frac{\Delta\tilde{k}}{2}(K-\bar{K})+B_\Delta\right)_s=0. \label{eta-s}
\end{align}
From this equation, we obtain
\begin{align}
|\mathcal{T}_s|^2\equiv \left|\mathcal{T}\right|^2\Big|_{\eta_s}\sim 1. \label{Ts=1}
\end{align}
Further, one can also verify $\left(\frac{d^2}{d^2\eta}|\mathcal{T}|^2\right)_s=-2|\dot{f}|^2_s\,\,(\dot{f}=\frac{df}{d\eta})$, which implies that the $\eta_s$ is an unstable point of $|\mathcal{T}|^2$ unless $\dot{f}_s=0$. Since the transmission coefficient decreases as the $\eta$ increases across the $\eta_s$ for $\dot{f}_s\neq 0$, the behavior of $|\mathcal{T}|^2$ near $\eta_s$ should be discussed carefully. In what follows, we treat $(K,\bar{K})$ under some approximations, where we will discuss the instability problem individually. Taking those sensitive problems into account, we use Eq.(\ref{transmission}) within the range $0<\eta\lesssim \eta_s$ tentatively.

Now, using the Eq.(\ref{eta-s}) for the denominator in the logarithmic function of the Eq.(\ref{ReE-1}), we obtain
\begin{align}
\Re\tilde{E} \sim \frac{1}{2\pi}\log \frac{1}{|\mathcal{T}_s|^2}\frac{\left(\eta_s-\frac{\Delta\tilde{k}}{2}\right)^2\left|(K+\bar{K})^2+2A_\delta\right|}{\left|\left(\eta_s+\frac{\Delta\tilde{k}}{2}\right)(K-\bar{K})-2B_\Delta\right|^2}. \label{ReE-2}
\end{align}
Here, we have left the $|\mathcal{T}_s|^2$ as a parameter without applying the Eq.(\ref{Ts=1}), since we sometimes use $(K,\bar{K})_s$ under some approximations; then, the Eq.(\ref{Ts=1}) does not exactly hold. Even so, for $|\eta_s(K-\bar{K})|\gg |2B_\Delta|_s$ or for the case of symmetric potential with $\delta=0$, the Eq.(\ref{ReE-2}) takes a simpler form such that
\begin{align}
\Re\tilde{E} \sim \frac{1}{2\pi}\log\frac{1}{|\mathcal{T}_s|^2}\left|\frac{K+\bar{K}}{K-\bar{K}}\right|^2. \label{ReE-3}
\end{align}
The Eq.(\ref{ReE-2}) or Eq.(\ref{ReE-3}) give the $\Re\tilde{E}$ as a function of $(\bar{K},\bar{K})_s$. Since the $(K,\bar{K})_s$ are again functions of $\tilde{E}$, the $\Re\tilde{E}$ under each approximation should be determined through those equations in a self-consistent manner.
\section{The asymptotic and the semi-classical approximations against $(K,\bar{K})$}

To know more nature of $(\mathcal{T},\mathcal{R})$, we need concrete forms of $(K,\bar{K})$. For this purpose, we consider two cases: i) an asymptotic approximation and ii) a semi-classical approximation to $(K,\bar{K})$. \\

\noindent
{\rm i}) The asymptotic region of $\phi_{(n)}(x)$ is defined by $\tilde{x}\gg \frac{1}{2}\sqrt{n(n-1)}$, where those states can be approximated by the asymptotic form given in Eq.(D\ref{asymptotic behavior}). The $\phi_{(n)}(x_b)$ takes such an asymptotic form depending on the $n$ and $(a_L,b_L)$ of $V_{RW}$; however, for $n=0,1$ states, one can apply the asymptotic form safely to them. Using the asymptotic form of $\phi_{(n)}(x_b)$, we can also define the asymptotic approximation of $\phi_{(n_E)}(x_b)$ by means of the analytic continuation $n\rightarrow n_E=n+i\Im{n},\,(n=\Re{n}\in \mathbb{N}_0)$. Then, by taking $\tilde{E}=-i\left(n_E+\frac{1}{2}\right)$ into account, we may use the $\Im\tilde{E}$ within $|\Im\tilde{E}|=\frac{1}{2},\frac{3}{2}$ safely .

Now, in the asymptotic region of $(\phi_{(n_E)}(x_b),\bar{\phi}_{(\bar{n}_E)}(x_b))$, the Eq.(D\ref{K-asymptotic}) gives rise to
\begin{align}
K \sim \left(\tilde{x}_b+\frac{\Re\tilde{E}}{\tilde{x}_b}\right)+\frac{i}{\tilde{x_b}}\left(\Im\tilde{E}+\frac{1}{2}\right), \label{asymp-1} \\
\bar{K} \sim \left(\tilde{x}_b+\frac{\Re\tilde{E}}{\tilde{x}_b}\right)+\frac{i}{\tilde{x_b}}\left(\Im\tilde{E}-\frac{1}{2}\right). \label{asymp-2}
\end{align}
When we apply those representations to Eq.(\ref{T-PNC}) and Eq.(\ref{ReE-2}), the asymptotic condition requires $\left(\frac{n}{\tilde{x_b}}\right)^2\sim\left(\frac{\Im\tilde{E}}{\tilde{x_b}}\right)^2\simeq 0$ so that $K+\bar{K}\sim 2\left(\tilde{x}_b+\frac{1}{\tilde{x}_b}\tilde{E}\right)$, $K-\bar{K}\sim \frac{i}{\tilde{x}_b}$, and $K\bar{K}\sim \left(\tilde{x}_b+\frac{\tilde{E}}{\tilde{x}_b}\right)^2+\frac{1}{(2\tilde{x}_b)^2}$. Thus, the Eq.(\ref{ReE-2}) is reduced to the Eq.(\ref{ReE-3}) under the conditions $\eta_s\gg \frac{1}{2}|\Delta\tilde{k}|$ and $\eta_s/\tilde{x}_b\gg2|B_\Delta|$; then, one can write the Eq.(\ref{ReE-3})
as
\begin{align}
\Re\tilde{E} & \sim \frac{1}{2\pi}\log \frac{2^2}{|\mathcal{T}_s|_{\rm as}^2}\left\{\left(\tilde{x}_b^2+\Re\tilde{E}\right)^2+\Im\tilde{E}^2\right\}. \label{ReE-app}
\end{align}
Further, one can verify $\dot{f}_s=-2i\tilde{E}/\tilde{x}_b$ from Eqs.(\ref{asymp-1})-(\ref{asymp-2}), the $\dot{f}_s$ is negligible under this approximation, and we don't need to worry about the instability at $\eta_s$.

The Eq.(\ref{ReE-app}) has a form $\Re\tilde{E}\sim F(\Re\tilde{E})$, where the function $F$ is suppressed by the factor $1/(2\pi)$. Then, setting $\Re\tilde{E}=0$ as a leading solution of $\Re\tilde{E}$, the $F(\Re\tilde{E}=0)$ gives the next leading solution in the sense of successive approximation for a smaller value of $|\Re\tilde{E}|$. Namely, in consideration of $\Im\tilde{E}=-\left(n+\frac{1}{2}\right)\,(n=0,1)$, the Eq.(\ref{ReE-app}) gives the approximate solution so that
\begin{align}
\Re\tilde{E} &\simeq \frac{1}{2\pi}\log \frac{2^2}{|\mathcal{T}_s|_{\rm as}^2}\left\{\tilde{x}_b^2+\Im\tilde{E}^2\right\} \label{ReE-s},\\
\Im\tilde{E} &=-\left(n+\frac{1}{2}\right).
\end{align}

On the other side, if we apply the Eq.(\ref{ReE-app}) to a larger $|\Re\tilde{E}|$, the Eq.(\ref{ReE-app}) becomes $\Re\tilde{E}\sim \frac{1}{2\pi}\log(2\Re\tilde{E})^2/|\mathcal{T}_s|^2_{\rm as}$, which can be solved formally using the Lambert $W$ function
\footnote{The Lambert W function $W(z)$ is the inverse function of $We^W=z$, which has two real branches of solution such as $W_0(z)>-1\,(z\in [-e^{-1},\infty])$ and $W_{-1}(z)<-1\,(z\in [-e^{-1},0])$.}
so that $\Re\tilde{E}\sim -\frac{1}{\pi}W_{-1}(-\frac{\pi}{2}|\mathcal{T}_s|_{\rm as})$. By definition of the Lambert function, the range of $W_{-1}(x)$ requires $|\mathcal{T}_s|_{\rm as}<\frac{2}{\pi e}\,(\simeq 0.23)$, which dissociates from the Eq.(\ref{Ts=1}). In other words, the case of $|\Re\tilde{E}|\gg 1$ is out of the application of this approximation. \\

\noindent
ii)~~ Another interesting way of studying $(K,\bar{K})$ is to use the semi-classical approximation, which is a counter-example of (i) that applies to a higher $n$ quantum number states in the region II. In this framework, by taking $\bar{K}_{(\bar{n}_E)_c}=\left(K_{(n_{E^*})c}\right)^*$ into account, one can derive that
\footnote{In the semi-classical approximation, it is understood that \\ $\phi_{(n)}\sim e^{\frac{i}{\hbar}\left(S_{(n)}^{(0)}+\frac{\hbar}{i}\tilde{S}_{(n)}^{(1)}\right)}$, $\left(\partial_{\tilde{x}}\tilde{S}_{(n)}^{(0)}\right)^2={2(\tilde{E}-\tilde{V}_M)}$, and $\tilde{S}_{(n)}^{(1)}=-\frac{1}{2} \partial_{\tilde{x}}\log\tilde{S}_{(n)}^{(0)}+\mbox{const.}\,(\tilde{S}^{(0)}=\frac{1}{\hbar}S^{(0)},x\in {\rm I\!I})$ under the condition\\ $\left|\tilde{x}\left(2(\tilde{E}-\tilde{V}_M\right)^{-\frac{3}{2}}\right|\ll 1$; then, $K=\partial_{\tilde{x}}\tilde{S}^{(0)}+\frac{i}{2}\partial_{\tilde{x}}\log\left(\partial_{\tilde{x}}\tilde{S}^{(0)}\right)$.
}

\begin{align}
K_{(n_E)c}&\sim \sqrt{2\left(\tilde{E}-\tilde{V}_M\right)_c}+\frac{i}{4}\frac{\tilde{x}_c}{\left(\tilde{E}-\tilde{V}_M\right)_c}, \label{semi-c-K} \\
\bar{K}_{(\bar{n}_E)c}&\sim \sqrt{2\left(\tilde{E}-\tilde{V}_M\right)_c}-\frac{i}{4}\frac{\tilde{x}_c}{\left(\tilde{E}-\tilde{V}_M\right)_c} \label{semi-c-bK}
\end{align}
, where $x_c \in {\rm I\!I}$. By taking $V_M(x_b)=0$ into account, these representations of $(K,\bar{K})$ derive the combinations at $\tilde{x}_b$ in such a form as $(K+\bar{K})_b\sim 2\sqrt{2\tilde{E}}, \, (K-\bar{K})_b\sim i\frac{\tilde{x}_b}{2\tilde{E}}$, and $(K\bar{K})_b\sim 2\tilde{E}+\left(\frac{\tilde{x}_b}{4\tilde{E}}\right)^2$.

By virtue of these equations, the Eq.(\ref{eta-s}) can be written as $|\eta_s^2-2\tilde{E}+i\frac{\Delta\tilde{k}}{2}\frac{\tilde{x}_b}{2\tilde{E}}-(B_{\Delta})_b|\sim 0$; that is, $\eta_s^2\sim 2|\tilde{E}|$ for a larger $|\tilde{E}|$. Since, then, $|(K-\bar{K})\eta_s|\sim \tilde{x}_b/\sqrt{2|\tilde{E}|}$, which decreases according to the $|\tilde{E}|$ increases, we must be careful about the presence or absence of $\delta\neq 0$. In the case of $(B_\Delta)_b\,(\propto \delta)\neq 0$, the Eq.(\ref{ReE-2}) becomes
\begin{align}
\Re\tilde{E}\sim \frac{1}{2\pi}\log \frac{1}{|\mathcal{T}_s|^2_{\rm s.c}}\frac{2|\tilde{E}|\left(2\sqrt{2|\tilde{E}|}\right)^2}{\left|i\frac{\tilde{x}_b}{\sqrt{2\tilde{E}}}-i\delta\left(2^3\tilde{E}+\frac{2\tilde{x}_b}{\tilde{E}}\right)\right|^2} \label{ReE-Delta}
\end{align}
, where we have used the relation $(B_{\Delta})_b\simeq \frac{i\delta}{2}\left(2^3\tilde{E}+\frac{2\tilde{x}_b}{\tilde{E}}\right)$ (Appendix E). From the Eq.(\ref{ReE-Delta}), we can see the following: \\

First, in the case of symmetric potential with $\delta=0$, the Eq.(\ref{ReE-3}) holds independent of the scale of $\eta_s$; and, we obtain
\begin{align}
\Re\tilde{E} \sim \frac{1}{2\pi}\log\frac{1}{|\mathcal{T}_s|^2_{\rm s.c}}\frac{2^5|\tilde{E}|^3}{\tilde{x}_b^2}. \label{ReE-semi-c}
\end{align}
Then for a smaller $|\Re\tilde{E}|\,\left(\ll |\Im\tilde{E}|\right)$, the successive approximation of $\Re\tilde{E}$ starting from the leading approximation $\Re\tilde{E}=0$ in the right-hand side of the Eq.(\ref{ReE-semi-c}) yields the counter form of the Eq.(\ref{ReE-s}) so that
\begin{align}
\Re\tilde{E} &\sim \frac{1}{2\pi}\log\frac{2^5}{|\mathcal{T}_s|^2_{\rm s.c}}\frac{1}{\tilde{x}_b^2}\left|\Im\tilde{E}\right|^3 \label{ReE-ss}
\end{align}
, which is again independent of the $\eta_s$. In this case, the quantum number $n$ is not restricted to $n=0,1$. On the contrarily for a larger $|\Re\tilde{E}|$, we may regard $|\tilde{E}|\sim \Re\tilde{E}$; and, the Eq.(\ref{ReE-semi-c}) is solved formally using the Lambert W function as $\Re\tilde{E}=-\frac{3}{2\pi}W_{-1}\left[-\frac{2\pi}{3}\left(\frac{|\mathcal{T}_s|^2_{\rm s.c}\tilde{x}_b^2}{2^5}\right)^{\frac{1}{3}}\right]$. Then the range of $W_{-1}$ requires $|\mathcal{T}_s|^2_{\rm s.c}<\frac{2^5}{\tilde{x}^2_b}\left(\frac{3}{2\pi e}\right)^3$$\,(\simeq 0.17/\tilde{x}_b^2)$, which is rather severe boundary condition to get $|\mathcal{T}_s|^2_{\rm s.c} \lesssim 1$.\\

Next, in the case of $\delta\neq 0$, the denominator in the right-hand side of Eq.(\ref{ReE-Delta}) comes to be dominated by the $\delta 2^3\tilde{E}$ for a larger $|\tilde{E}|$ exceeding $\frac{1}{2}\left(\frac{\tilde{x}_b}{4\delta}\right)^{\frac{2}{3}}$; then,
\begin{align}
\Re\tilde{E} \sim \frac{1}{2\pi}\log\frac{1}{|\mathcal{T}_s|^2_{\rm s.c}2^2\delta^2}.
\end{align}
Since the right-hand side of this equation should be larger than $\frac{1}{2}\left(\frac{\tilde{x}_b}{4\delta}\right)^{\frac{2}{3}}$, we obtain $|\mathcal{T}_s|^2_{\rm s.c}\tilde{x}^2_b>(\delta/\tilde{x}_b)^{-2}e^{-\pi(4(\delta/\tilde{x}_b))^{-\frac{2}{3}}}$. The right-hand side of this inequality gives the lower bound of $|\mathcal{T}_s|^2_{\rm s.c}$ such as $0.7/\tilde{x}_b^2$. Since the upper bound of $|\mathcal{T}_s|^2_{\rm s.c}$ is free in this case, we may regard as $0.7/\tilde{x}_b^2<|\mathcal{T}_s|^2_{\rm s.c}\lesssim 1$.

Further, it is not difficult to derive $\dot{f}_s\sim \frac{i}{\tilde{x}_b}(\frac{\tilde{x}_b}{\tilde{E}})^2$ out of $(K\pm\bar{K},K\bar{K})$ in this approximation. Then the $\dot{f}_s$ tends to zero according as the $\eta$ increase beyond $\eta_s$, and there is no instability problem in this case too. \\

Now, the two approximations, the asymptotic approximation and the semi-classical approximation, are effective in the different regions of $(\Re\tilde{E},\Im\tilde{E})$. However, since both the Eq.(\ref{ReE-app}) and Eq.(\ref{ReE-semi-c}) are representations for a smaller $|\Re\tilde{E}|$ with negligible $O(\delta)$ terms, it may be interesting to write those representations in one form in the sense of interpolation. The way for this purpose is not unique, and a simple one is to put
\begin{align}
\begin{split}
\Re\tilde{E} &\sim \frac{1}{2\pi}\log\frac{2^2\tilde{x}_b^2}{|\mathcal{T}_s|^2_{\rm as}}\left\{1+\left(\frac{\Im\tilde{E}}{\tilde{x}_b}\right)^2+\frac{|\mathcal{T}_s|^2_{\rm as}2^3}{|\mathcal{T}_s|^2_{\rm s.c}\tilde{x}_b}\left(\frac{\Im\tilde{E}}{\tilde{x}_b}\right)^3\right\}, \\
\Im\tilde{E} &~=-\left(n+\frac{1}{2}\right)\,(n=0,1,2,\cdots) \label{interpolation}
\end{split}
\end{align}
, which covers the Eq.(\ref{ReE-s}) for a smaller $|\Im\tilde{E}|$ and the Eq.(\ref{ReE-ss}) for a larger $|\Im\tilde{E}|$ respectively. In particular, when the coefficient
\footnote{The $N$ should be decided depending on the parameters of potential. For example for the $V_{RW}$ with $(J,L)=(3,2)$, one can verify $\frac{\tilde{x}_b}{2}\simeq 0.23$; and so, $|\mathcal{T}_s|^2_{\rm as}\sim 0.23 N|\mathcal{T}_s|^2_{\rm s.c}$. If we apply the Eq.(\ref{Ts=1}) to $|\mathcal{T}_s|^2$, then we must set as $N\lesssim (0.23)^{-1}$.}
$N\equiv |\mathcal{T}_s|^2_{\rm as}2^3/|\mathcal{T}_s|^2_{\rm s.c}\tilde{x}_b$ is a numerical constant, there arises an invariance for the Eq.(\ref{interpolation}) under the simultaneous transformations $\Im\tilde{E}\rightarrow \alpha\Im\tilde{E},\tilde{x}_b\rightarrow \alpha\tilde{x}_b$, and $\Re\tilde{E}\rightarrow \Re\tilde{E}+\frac{1}{2\pi}\log\alpha$. That is a scale transformation of $(\Im\tilde{E},\tilde{x}_b)$ is related to a logarithmic sift of the $\Re\tilde{E}$, which is consistent with the smallness of the $\Re\tilde{E}$.
Thus, the Eq.(\ref{interpolation}) should be read as a counter form of the Eq(\ref{frequency}).

\section{Summary and discussion}

In this paper, we have discussed the frequency structure of the QNM from the viewpoint of the scattering of particles interacting with the Regge-Wheeler potential $V_{RW}$, which characterizes the slight fluctuation of a black hole.
In terms of the tortoise coordinate, the wave equation for the QNM can be reduced to a one-dimensional Schr$\ddot{\rm o}$dinger equation with a potential having the shape of convex upward with asymptotic level lines on both sides.

We have approximated this potential simply by the $V_M$ consisting of the regions I, I\!I, and I\!I\!I. In the regions I and I\!I\!I, the $V_M$ takes constant values, where the states are set as plane waves with the wave number $k$ and $k^\prime$ respectively, for incoming and transmitting particles. In the region I\!I, the potential takes a truncated form of the parabolic $V_K$. The totality of wave functions under the $V_M$ is obtained by matching the wave functions in respective regions by the continuity of those functions and their derivative functions. However, the way of setting quantum states in the region I\!I is ambiguous. Under the potential $V_K$, the Schr$\ddot{\rm o}$dinger equation can be solved exactly, and $(\phi_{(n)},\bar{\phi}_{(n)})\, (n\in\mathbb{N}_0)$ are obtained as a complete basis. We have adopted a linear combination of those states as the wave function in this region
\footnote{The E-representation of Green's function, the resolvent $G_E(x)=\langle x|(E-H_r)^{-1}|x_c\rangle,\,(x_c\notin I\!I)$ is a possible candidate instead of the $\phi_n(x)$ in setting the $\psi_E^{(I\!I)}(x)$. However, under the semiclassical aproximation, it holds that $G_E(x)\propto \phi_{(n_E)}(x)\phi_{(n_E)}(x_c)$; and so, the same representation of $(\mathcal{T},\mathcal{R})$ is obtained in this case too. }
, since the solutions in existence suggest the energy spectrum associated with the states $(\phi_{(n)},\bar{\phi}_{(n)})$.

According to this approach to the scattering problem of particles under the potential $V_M$, we can obtain formal representations for the transmission and reflection coefficient densities, the $\mathcal{T}$ and $\mathcal{R}$. Furthermore, using the particle number conservation $|\mathcal{T}|^2+|\mathcal{R}|^2=1$, we could derive a functional equation among the particle energies $\tilde{E}=\Re\tilde{E}+i\Im\tilde{E}$, which are complex values in general, and the $\eta=\frac{1}{2}(\tilde{k}^\prime+\tilde{k})$. As a result, the transmission coefficient $|\mathcal{T}|^2$ is obtained as a monotonically increasing function of $\eta$ until its stationary point $\eta_s$. Further, in the neighborhood of $\eta_s$, the transmission coefficient $|\mathcal{T}_s|^2\equiv |\mathcal{T}|^2_{\eta=\eta_s}$ is found to be proportional to the factor $e^{-2\pi\Re\tilde{E}}$, which gives rise to a Hawking-like temperature for the transmitting particles. Another important point associated with the particle number conservation is that the $\Re\tilde{E}$, the counterpart of $\omega$ in the Eq.(\ref{frequency}), can be characterized as a solution of the functional equation including $(\Re\tilde{E},\Im\tilde{E},|\mathcal{T}_s|^2)$.

To analyze the $\Re\tilde{E}$ more concretely, we considered the two approximations, the asymptotic and semi-classical approximations. The asymptotic approximation is effective for a smaller $|\Im\tilde{E}|$; and the representation of the $\Re\tilde{E}$ based on the Eq.(\ref{ReE-2}) is also obtained safely for its smaller value.

On the other side, the semi-classical approximation is sensitive to the effect of $O(\delta)$ terms in the Eq.(\ref{ReE-2}) to evaluate $\Re\tilde{E}$. For a smaller value of $|\Re\tilde{E}|$, both approximations can yield respective solutions effective in different regions of $|\Im\tilde{E}|$. To study the meaning of those solutions in various scales of $|\Im\tilde{E}|$, we first treated the transmission coefficients in respective solutions, the $|\mathcal{T}_s|^2_{as}$ and $|\mathcal{T}_s|^2_{s.c}$, as parameters associated with those approximations. Then, we integrated those solutions under different approximations into one form using the interpolation. In concrete terms, we added two solutions. As a result, the integrated form applies to a smaller $|\Re\tilde{E}|$ with any value of $|\Im\tilde{E}|$, which should be compared with the Eq.(\ref{frequency}) in the present way.

Now, in the integration of two solutions, the factor $N=|\mathcal{T}_s|^2_{\rm as}2^3/|\mathcal{T}_s|^2_{\rm s.c}\tilde{x}_b$ plays a role of a constant, which controls a symmetry between two original solutions. In particular for $N=1$, we obtain $|\mathcal{T}_s|^2_{\rm as}\lesssim \frac{\tilde{x}_b}{2^3}|\mathcal{T}_s|^2_{\rm s.c}$; that is, the transmission coefficient under the asymptotic approximation is smaller than one under the semi-classical approximation.

Furthermore, under the setting of $N=1$, there arises a symmetry of the integrated representation for the $|\Re\tilde{E}|$ under the simultaneous transformations $\Im\tilde{E}\rightarrow \alpha\Im\tilde{E},\tilde{x}_b\rightarrow \alpha\tilde{x}_b$, and $\Re\tilde{E}\rightarrow \Re\tilde{E}+\frac{1}{2\pi}\log\alpha$. If the integrated representation is a counterpart of the Eq.(\ref{frequency}), the above scale transformations will change its leading logarithmic constant. More investigation for such a scale dependence of $\Re\tilde{E}$ will be important problems in future work.

\section*{Acknowledgments}

The authors thank the College of Science and Technology, Nihon University, for supporting the research.

\appendix
\section{The parameters characterizing $V_{RW}(x)$}

Since $r^4V_{RW}(r)$ is a second degree polynomial in $r$, one can factorize $V_{RW}(x)$ in terms of dimensionless variable $\tilde{r}=\frac{r}{r_H}$ so that
\begin{align}
\begin{split}
V_{RW}(r) &=\frac{1}{r_H^2}\frac{a_L}{\tilde{r}^4}\Big(\tilde{r}-v_{+}\Big)\Big(\tilde{r}-v_{-}\Big) \\
&=\frac{1}{r_H^2}\tilde{V}_{RW}(\tilde{r})
\end{split}
\end{align}
, where $v_{+}+v_{-}=\frac{a_L+b_J}{a_L}$ and $v_{+}v_{-}=\frac{b_J}{a_L}$; i.e.,
\begin{align}
v_\pm &=\frac{(a_L+b_J) \pm (a_L-b_J)}{2a_L}
\end{align}
for $a_L\geq b_J$. The $v_{\pm}$ are zeros of $V_{RW}(r)$; for example for $(L,J)=(3,2)$, the case $v_{+}=1$(horizon) reflect to $v_{-}=\frac{b_J}{a_L}=\frac{1}{4}$(inside horizon). Further, by taking $\partial_xV_{RW}(r)=f(r)\partial_rV_{RW}(r)$ into account, we obtain with $\tilde{x}=\frac{x}{r_H}$,
\begin{align}
\partial_{\tilde{x}}\tilde{V}_{RW}(\tilde{r})=-\frac{2a_L}{\tilde{r}}f(\tilde{r})\left(\tilde{r}-\tilde{r}_{+}\right)\left(\tilde{r}-\tilde{r}_{-}\right)
\end{align}
, where $\tilde{r}_{\pm}$ are stationary points of $\tilde{V}_{RW}$ satisfying $\partial_{\tilde{x}}\tilde{V}_{RW}\big|_{\tilde{r}_\pm}=0$; those are related to $v_{\pm}$ by
\begin{align}
\tilde{r}_{+}+\tilde{r}_{-}=\frac{3}{2}\big(v_{+}+v_{-}\Big),~\tilde{r}_{+}\tilde{r}_{-}=2v_{+}v_{-}.
\end{align}
As for $(J,L)=(3,2)$, one can evaluate $\tilde{r}_{+}\simeq 1.55\cdots$ and $\tilde{r}_{-}<1$ (inside horizon); and so, only $\tilde{r}_{+}$ is important as a physical stationary point. Then one can also verify $\tilde{V}_{RW}(\tilde{r}_{+})\simeq 1.48\cdots$.

Similarly, one can derive the second order derivative of $\tilde{V}_{RW}(\tilde{r})$. We are interested in, however, the value at the stationary point $\tilde{r}_{+}(\tilde{x_{+}})$, to which we can obtain
\begin{align}
\partial^2_{\tilde{x}}\tilde{V}_{RW}(\tilde{r})\big|_{\tilde{r}_{+}}=-\frac{2a_{L}f(\tilde{r}_{+})^2}{\tilde{r}_{+}^4}\left(1-\frac{2b_J}{a_L\tilde{r_{+}}}\right).
\end{align}
In particular for the case of $(L,J)=(3,2)$ taken as an example, we obtain $\partial^2_{\tilde{x}}\tilde{V}_{RW}(\tilde{r})\big|_{\tilde{r}_{+}}\simeq -0.41\cdots$. The curve of parabolic approximation of $\tilde{V}_{RW}(\tilde{x})$ in Fig.1 has been plotted with those parameters.

We also point out that the $x_b=l_*\tilde{x}_b$ is a point satisfying $V_K(x_b)=\hbar\omega_*\left(\tilde{V}_0-\frac{1}{2}\tilde{x}_b^2\right)=0\, \left(\tilde{V}_0=\frac{1}{\hbar\omega_*}V_0\right)$; that is, $\tilde{x}_b=\sqrt{2\tilde{V}_0}$. The simple manipulation leads to
\begin{align}
\tilde{x}_b=\sqrt{\frac{(\tilde{V}_{RW})_M}{\sqrt{\frac{1}{2}\left|\partial^2_{\tilde{x}}\tilde{V}_{RW}\right|_M}}} \label{x_b}
\end{align}
, where the suffix $M$ means the values at the maximal point of $V_{RW}$. In the case of $(L,J)=(3,2)$, for example, we can evaluate from this equation so that $\tilde{x}_b\simeq 1.80$.

\section[Green]{Green's function for $\hat{H}_r$}

To get the expression (\ref{Green-T2}), we start with
\begin{align}
G(T;x_x,x_b) &\equiv \sum_{n=0}^\infty \frac{1}{N_n}e^{-\omega_*\left(n+\frac{1}{2}\right)T}\phi_{(n)}(x_b)\bar{\phi}_{(n)}(x_a)^* \nonumber \\
=e^{-\frac{\omega_*}{2}T} &\sqrt{\frac{2\pi}{i}}\sqrt{\frac{m_*\omega_*}{2\hbar\pi^2}}e^{e^{-\left(\frac{i\pi}{2}+\omega_* T\right)}\bar{A}_b\bar{A}_a} e^{i\frac{m_*\omega_*}{2\hbar}(x_a^2+x_b^2)} \label{G-T}
\end{align}
, where the use has been made of Eqs. (\ref{phi-n}) and (\ref{E-n}) in addition to $\bar{\phi}_{(n)}=\phi_{(n)}^*$ in the $x$ representation. The Eq.(\ref{ladder}) with $\alpha(T)=e^{-\left(\frac{i\pi}{2}+\omega_* T\right)}$ gives rise to
\begin{align}
e^{\alpha(T)\bar{A}_b\bar{A}_a} &=e^{\alpha(T)\sqrt{\frac{m_*\omega_*}{2\hbar}}x_bA_a^*}\nonumber \\
&\times e^{\alpha(T)\frac{1}{\sqrt{2m_*\hbar\omega_*}}\hat{p}_b\bar{A}_a}e^{-\frac{i}{4}\alpha(T)^2(\bar{A}_a)^2}. \label{formula}
\end{align}
Here, the repeated use of the Gaussian integral
\begin{align}
e^{\frac{i}{2}M^2}=\sqrt{\frac{i}{2\pi}}\int_{-\infty}^\infty dk
e^{-\frac{i}{2}k^2+ikM}
\end{align}
allows us to write
\begin{align}
&e^{\alpha(T)\bar{A}_bA_a^*}e^{i\frac{m_*\omega_*}{2\hbar}(x_a^2+x_b^2)} \nonumber \\
&=e^{i\frac{m_*\omega_*}{2\hbar}x_b^2}\sqrt{\frac{i}{2\pi}}\!\int_{-\infty}^\infty\!\!\! dke^{-\frac{i}{2}k^2}e^{i\left(k-2\sqrt{\frac{m_*\omega_*}{2\hbar}}x_b\right)e^{-\omega_* T}\sqrt{\frac{m_*\omega_*}{2\hbar}}x_a} \nonumber \\
&=e^{-i\frac{m_*\omega_*}{2\hbar}x_b^2}\frac{1}{\sqrt{1-e^{-2\omega_* T}}}
e^{i\frac{m_*\omega_*}{\hbar}\frac{(e^{-\omega_* T}x_a-x_b)^2}{(1-e^{-2\omega_* T})}}e^{i\frac{m_*\omega_*}{2\hbar}x_a^2}. \label{little long}
\end{align}
Then, it is a simple exercise to reduce the Eq.(\ref{G-T}) with the Eq.(\ref{little long}) to the Eq.(\ref{Green-T2}).

\section{Analytic continuation of derivative functions}

By definition of derivative function $\phi_{(n_E)\mathring{a}}$, Eq.(\ref{a-a_*}) gives
\begin{align}
\phi_{(n_E)\mathring{a}}
&= \lim_{\epsilon\rightarrow 0}\frac{e^{i\pi n_E}}{\epsilon}\left(\phi_{(n_E)a_*-\epsilon}-\phi_{(n_E)a_*}\right) \nonumber \\
&=-e^{i\pi n_E}\phi_{(n_E)\mathring{a}_*} \, \label{a-dot}
\end{align}
by taking the analytic continuation of two points $(\tilde{x}_{a+\epsilon},\tilde{x}_a)$ $=e^{i\pi}(\tilde{x}_{a_*-\epsilon},\tilde{x}_{a_*})$ into account; and, the Eq.(\ref{bar-K-dot}) is a result of the complex conjugate of this equation. Furthermore, remembering $\bar{n}_E\equiv \left(n_{E^*}\right)^*=-i\tilde{E}-\frac{1}{2}=-n_E-1$, we obtain
\begin{align}
\bar{\phi}_{(\bar{n}_E)\mathring{a}} &=\left(\phi_{(n_{E^*})\mathring{a}}\right)^*=\left(-e^{i\pi n_{E^*}}\phi_{(n_{E^*})\mathring{a}_*}\right)^* \nonumber \\
&=-e^{-i\pi\bar{n}_E}\left(iK_{(n_{E^*})a_*}\phi_{(n_{E^*})a_*}\right)^* \nonumber \\
&=-e^{i\pi n_E}i\bar{K}_{(\bar{n}_E)a_*}\bar{\phi}_{(\bar{n}_E)a_*} \label{bar-a-dot}
\end{align}
with the notation $\bar{K}_{(\bar{n}_E)a_*}=\left(K_{(n_{E^*})a_*}\right)^*$. Equations (\ref{a-dot}) and (\ref{bar-a-dot}) check the validity of Eqs. (\ref{K-dot}) and (\ref{bar-K-dot}) respectively.

Now, on the order of smallness, we had set $\Delta\tilde{k}^2\sim \delta$ and $\delta^2\simeq 0$ as discussed in section 2. In such a non-symmetric $(\Delta\tilde{k}\neq 0)$ case, the forms of $(\mathcal{T},\mathcal{R})$ are not so complex. Indeed, within the first order of $\delta$, we obtain
\begin{align}
\phi_{(n_E)a_*} &\simeq e^{i\delta K_{(n_E)b}}\phi_{(n_E)b}, \\
\phi_{(n_E)\mathring{a}_*} &=\lim_{\epsilon\rightarrow 0}\frac{1}{\epsilon}\left(\phi_{(n_E)b+\delta+\epsilon}-\phi_{(n_E)b+\delta}\right) \nonumber \\
&\simeq \lim_{\epsilon\rightarrow 0}\frac{1}{\epsilon}\left[\left\{\phi_{(n_E)b}+(\delta+\epsilon)\phi_{(n_E)\mathring{b}} \right.\right. \nonumber \\
&\left.\left.+\frac{(\delta+\epsilon)^2}{2!}\partial_b\phi_{(n_E)\mathring{b}}\right\} \right. -\left\{\phi_{(n_E)b}+\delta\phi_{(n_E)\mathring{b}} \right\}\Bigg] \nonumber \\
&\simeq\left\{1+\delta\left(iK^{(n_E)}_b+L^{(n_E)}_b\right)\right\}\phi_{(n_E)\mathring{b}} \nonumber \\
&\simeq e^{i\delta\left(K_{(n_E)b}-iL_{(n_E)b}\right)}\phi_{(n_E)\mathring{b}} \label{a*-dot}
\end{align}
, where $L_{(n_E)_b}=\left(\log K_{(n_E)}\right)_{\mathring{b}}=K_{(n_E)\mathring{b}}/K_{(n_E)b}$. Thus, in consideration of Eqs.(\ref{a-a_*})-(\ref{bar-K-dot}), the above ways associating the states at $x_a$ to those at $x_b$ are able to give
\begin{align}
\begin{split}
\phi_{(n_E)a} &\simeq e^{i\pi n_E}e^{i\delta_a K_{(n_E)b}}\phi_{(n_E)b} , \\
\bar{\phi}_{(\bar{n}_E)a} &\simeq -e^{i\pi n_E}e^{-i\delta_a \bar{K}_{(\bar{n}_E)b}}\bar{\phi}_{(\bar{n}_E)b} , \\
\phi_{(n_E)\mathring{a}} &\simeq -e^{i\pi n_E}e^{i\delta_a K^L_{(n_E)b}}iK_{(n_E)b}\phi_{(n_E)b} , \\
\bar{\phi}_{(\bar{n}_E)\mathring{a}} &\simeq -e^{i\pi n_E}e^{-i\delta_a \bar{K}^L_{(\bar{n}_E)b}}i\bar{K}_{(\bar{n}_E)b}\bar{\phi}_{(\bar{n}_E)\mathring{b}}
\end{split}
\end{align}
with the notation $K^L_{(n_E)b}\equiv K_{(n_E)b}-iL_{(n_E)b}$ and $\bar{K}^L_{(n_E)b}=\left(K^L_{(n_{E^*})b}\right)^*$. Applying these equations to $(\Phi_{ab},\Phi_{\mathring{a}b},\Phi_{a\mathring{b}},\Phi_{\mathring{a}\mathring{b}})$ and $\Phi_{b\mathring{b}}$, we obtain the following:
\begin{align}
\begin{split}
\Phi_{ab} &\simeq \left\{e^{i\left(\pi n_E+\delta_a K_{(n_E)b}\right)}\right. \\
&\left. +e^{i\left(\pi n_E-\delta_a\bar{K}_{(\bar{n}_E)b}\right)}\right\}\phi_{(n_E)b}\bar{\phi}_{(\bar{n}_E)b}, \\
\Phi_{a\mathring{b}} &\simeq -i\left\{e^{i\left(\pi n_E+\delta_aK_{(n_E)b}\right)}\bar{K}_{(\bar{n}_E)b} \right. \\
&\left. -e^{i\left(\pi n_E-\delta_a\bar{K}_{(\bar{n}_E)b}\right)}K_{(n_E)b} \right\}\phi_{(n_E)b}\bar{\phi}_{(\bar{n}_E)b}, \\
\Phi_{a\mathring{b}} &\simeq -i\left\{e^{i\left(\pi n_E+\delta_aK_{(n_E)b}\right)}\bar{K}_{(\bar{n}_E)b} \right. \\
&\left. -e^{i\left(\pi n_E-\delta_a\bar{K}_{(\bar{n}_E)b}\right)}K_{(n_E)b} \right\}\phi_{(n_E)b}\bar{\phi}_{(\bar{n}_E)b}, \\
\Phi_{\mathring{a}\mathring{b}} &\simeq -\left\{e^{i\left(\pi n_E+\delta_a\langle K_{(n_E)b}\right)} \right. \\
&\left. +e^{i\left(\pi n_E-\delta_a\langle \bar{K}_{(\bar{n}_E)b}\rangle\right)}\right\} K_{(n_E)b}\bar{K}_{(\bar{n}_E)b}\phi_{(n_E)b}\bar{\phi}_{(\bar{n}_E)b},
\end{split}
\end{align}
and
\begin{align}
\Phi_{b\mathring{b}} &\simeq -i\left(\bar{K}_{(\bar{n}_E)b}+K_{(n_E)b}\right)\phi_{(n_E)b}\bar{\phi}_{(\bar{n}_E)b} \,.
\end{align}
With the aid of these equations, Eqs.(\ref{T-eq}) and (\ref{R-eq}) turn out to be Eq.(\ref{AT-eq}) and the Eq,(\ref{AR-eq}) respectively.

\section{Asymptotic behavior of $\phi_{(n)}(x)$ and related matter} \label{asymptotic}

The asymptotic behavior of $\phi_{(n)}(x)$ can be derived easily from its integral representation. In terms of the dimensionless variable $\tilde{x}=l_*^{-1}x$, one can write that $\bar{A}=\frac{1}{\sqrt{2}}\left(\tilde{x}-i\partial_{\tilde{x}}\right)$ and $\phi_{(0)}\left(x(\tilde{x})\right)=N_*e^{\frac{i}{2}\tilde{x}^2} \left(N_*=\sqrt[4]{\frac{m_*\omega_*}{2\hbar\pi^2}}\right)$. Then due to Eq.(\ref{phi-n}) and $e^{z\bar{A}}=e^{-\frac{i}{4}z^2+\frac{z}{\sqrt{2}}\tilde{x}}e^{-i\frac{z}{\sqrt{2}}\partial_{\tilde{x}}}$, we obtain the expression
\begin{align}
\phi_{(n)}\left(x(\tilde{x})\right) &=N_*\frac{n!}{2\pi i}\int_{C}\frac{dz}{z^{n+1}}e^{z\bar{A}}e^{\frac{i}{2}\tilde{x}^2} \nonumber \\
&=N_*\frac{n!}{2\pi i}\int_{C}\frac{dz}{z^{n+1}}e^{\frac{i}{2}\tilde{x}^2-\frac{i}{2}z^2+\sqrt{2}\tilde{x} z} \label{integral-rep}
\end{align}
for a positive integer $n$. Here $C$ is a closed path going around counterclockwise $z=0$. The Eq.(\ref{integral-rep}) allows us the analytic continuation with respect to $n$ through the equation $n!/z^n=\Gamma(n+1)e^{-n\log z}$. Further, under the scale transformation $z=\frac{\sqrt{2}}{\tilde{x}}z^\prime$, the following expansion holds:
\begin{align*}
\phi_{(n)}\left(\tilde{x}\right) &=N_*\frac{n!}{2\pi i}\tilde{x}^n e^{\frac{i}{2}{\tilde{x}^2}}\oint_{C^\prime}\frac{dz^\prime}{z^{\prime n+1}}e^{-i\left(\frac{z^\prime}{\eta}\right)^2+2z^\prime} \\
&=\frac{N_*}{\sqrt{2^n}}\frac{n!}{2\pi i}\tilde{x}^n e^{\frac{i}{2}\tilde{x}^2}\times 2\pi i\left\{\frac{2^n}{n!}-\frac{i}{\tilde{x}^2}\frac{2^{n-2}}{(n-2)!}+\cdots \right\} \\
&=N_*\sqrt{2^n}\tilde{x}^n e^{\frac{i}{2}\tilde{x}^2}\left(1-\frac{i}{\tilde{x}^2}\frac{n(n-1)}{2^2}\cdots\right).
\end{align*}
Therefore, the well-known asymptotic behavior
\begin{align}
\phi_{(n)}(\tilde{x})\equiv \phi_{(n)}\left(x(\tilde{x})\right)\approx N_*\sqrt{2^n}\tilde{x}^n e^{\frac{i}{2}\tilde{x}^2} \label{asymptotic behavior}
\end{align}
is derived under the condition $\big|\tilde{x}\big|\gg \frac{1}{2}\sqrt{n(n-1)}$,
which is satisfied automatically for $n=0,1$; on the other side, the satisfaction of that condition for $n \geq 2$ is dependent on $(a_L,b_J)$ of $V_{RW}$.
Further, the logarithmic derivative of $\phi_{(n)}(\tilde{x}_c)$ in Eq.(\ref{asymptotic behavior}) yields
\begin{align}
K_{(n)c} &\equiv -i\partial_{\tilde{x}_c}\log \phi_{(n)c}\approx \tilde{x}_c-i\frac{n}{\tilde{x}_c}. \label{K-asymptotic}
\end{align}
in the asymptotic region of $\phi_{(n)c}$. However, it is not difficult to derive directly $K_{(n)c}\approx i\tilde{x}_c\phi_{(n)c}$ and $K_{(n)c}\approx \frac{n}{\tilde{x}_c}\phi_{(n)c}\phi_{(n)c}$ respectively for $|\tilde{x}_c|\gg n$ and $|\tilde{x}_c|\ll n$ from the integral representation (\ref{integral-rep}).
We finally note that the states $\{\phi_{(n_E)}\}$ in section 6 are those defined by the substitution of $n\rightarrow n_E=n+i\Im n_E \, (n=\Re{n}\in \mathbb{N}_0)$ in the Eq.(\ref{asymptotic behavior}); and so, those asymptotic forms are available to use safety for $\Re{n}_E=0,1$.

\section{The forms of $A_\delta$ and $B_\Delta$ under the approximations \lq asymptotic\rq ~and \lq semi-classical\rq}
As for the use in section 6, we here summarize the forms of $(A_\delta, B_\Delta)$, the Eq.(\ref{A-delta}) and Eq.(\ref{B-Delta}), under the asymptotic and semi-classical approximations. In this case, it is sufficient to know the combinations $K+\bar{K},K-\bar{K}$, and $K\bar{K}$.\\

\noindent
{\bf The asymptotic approximation:} In this case, by virtue of the (\ref{asymp-1}) and (\ref{asymp-2}), one can obtain
\begin{align}
K+\bar{K} &\sim 2\left\{\left(\tilde{x}_b+\frac{1}{\tilde{x}_b}\Re{\tilde{E}}\right)+\frac{i}{\tilde{x}_b}\Im{\tilde{E}}\right\}, \label{K+Kbar} \\
K-\bar{K} &\sim \frac{i}{\tilde{x}_b}, \\
K\bar{K} &\sim \left(\tilde{x}_b^2+2\Re{\tilde{E}}\right)+2i\Im{\tilde{E}} \label{KKbar}
\end{align}
, from which the following follows:
\begin{align}
A_\delta 
&=\delta\left(\frac{1}{\tilde{x}_b^3}+\frac{1}{\tilde{x}_b^2}\right) , \\
B_\Delta 
&\simeq 2i\delta \left[\left\{\left(\tilde{x}_b+\frac{1}{\tilde{x}_b}\Re{\tilde{E}}\right)+\frac{i}{\tilde{x}_b}\Im{\tilde{E}}\right\}^2+\frac{1}{\tilde{x}_b}\right]
\end{align}
disregarding the term of the order $O(\Delta\tilde{k}^2)$. \\

\noindent
{\bf The semi-classical approximation:} From equations (\ref{semi-c-K}) and (\ref{semi-c-bK}) at $\tilde{x}_b$, it can be verified that
\begin{align}
\left(K+\bar{K}\right)_b &\sim 2\sqrt{2\tilde{E}} , \label{semi+} \\
\left(K-\bar{K}\right)_b &\sim \frac{i}{2}\frac{\tilde{x}_b}{\tilde{E}}, \\
\left(K\bar{K}\right)_c &\sim 2\tilde{E}+\frac{1}{16}\frac{\tilde{x}^2_c}{\tilde{E}^2} \label{semi-KKbar}
\end{align}
because of $V_M(\tilde{x}_b)=0$. Then substituting those combinations for the Eq.(\ref{A-delta}) and Eq.(\ref{B-Delta}), one can derive
\begin{align}
A_\delta &=\delta\left\{\left(\frac{1}{2}\frac{\tilde{x}_b}{\tilde{E}}\right)^3+\left(\frac{1}{2}\frac{\tilde{x}_b}{\tilde{E}}\right)^2\right\}, \\
B_\Delta &\simeq i\delta\left\{\frac{1}{2}\left(2\sqrt{2\tilde{E}}\right)^2+\frac{\tilde{x}_b}{\tilde{E}}\right\}
\end{align}
, where the $O(\Delta\tilde{k}^2)$ term is again discarded.

%

%
%
%
%

\end{document}